\documentclass[%
 aip,
 jmp,%
 amsmath,amssymb,
 reprint,%
]{revtex4-1}

\usepackage{graphicx}
\usepackage{subfigure}
\usepackage{dcolumn}
\usepackage{bm}

\begin{document}

\preprint{AIP/123-QED}

\title[Sample title]{Pocket doping method on Improving the Performance of Graphene Nanoribbons Tunneling FET }

%
\author{Fei Liu, Xiaoyan Liu, Jinfeng Kang, Yi Wang}
\affiliation{
Institute of Microelectronics,Peking University, Beijing 100871, People¡¯s Republic of China
}%

\date{Dated: 23 November 2012}

\begin{abstract}
A tunneling field effect transistor based on armchair graphene nanoribbons is studied using ballistic quantum transport simulation based on 3D real space nonequilibrium Green's function formalism. By introducing a pocket doping region near the source, the performance of GNRs TFET is improved in terms of larger on current and steeper subthreshold swings compared to conventional tunneling FET. It is also found that pocket region introduces two interesting features. By increasing the pocket length the ambipolar can be effectively suppressed. Furthermore, there is negative trans-conductance at negative gate voltage due to resonance.
\end{abstract}

\maketitle
Tunneling field-effect transistors(TFET) have been  the topic of studies in recent years due to excellent properties  compared to conventional semiconductor devices , which  make it possible to reduce power dissipation in integrated circuits\cite{1,2}. In a classical Metal-Oxide-Semiconductor device, the device state is controlled by modulating thermionic current over an energy barrier. While, TFET has a different mechanics using quantum mechanical tunneling. Relying on the tunneling current, TFET can overcome the fundamental limit to subthreshold swing less than 60 mV/decade at room temperature. At the same time, due to the tunneling resistance the on sate current is limited. Hence, various methods and materials are proposed to improve TFET performance \cite{3,4,5,6,7,8,9,10}. Especially, graphene related material is studied, due to its lighter carrier effective masses\cite{9,10}. It is expected TFET based on graphene can have high on state current. It has been shown that the performance of graphene nanoribbons (GNRs) TFET can be modulated by varying the ribbon width, which the band gap depends on. Furthermore, improved GNRs TFET design can be got by using strain methods or heterostructure\cite{12,13,14,15,16}.

 \begin{figure}
\includegraphics[width=3.2in,height=2.6in]{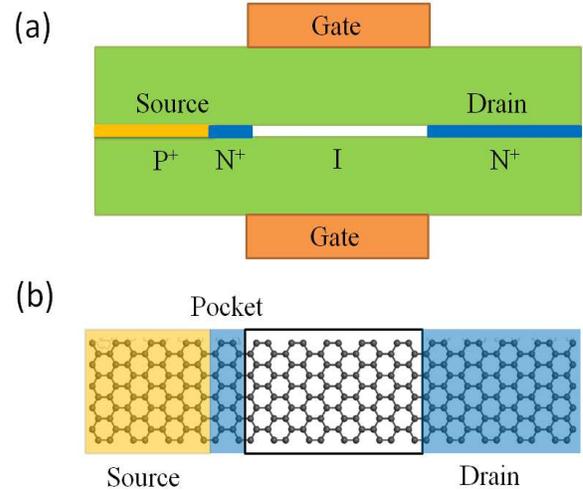}
\caption{(a) Schematic of simulated double gate GNRs TFETs  with  $\texttt{SiO}_2$ oxide thickness of tox = 1.0 nm  and dielectric constant of  k=4  (b) Atomistic structure of armchair GNRs used as transport material, which is composed of N=13 AGNR. The p-doped source and n-doped drain have the same doping density of 0.01 dopant/atom. A pocket region is proposed with doping density of 0.01 dopant/atom.}\label{fig1}
\end{figure}
In this letter, we theoretically study GNRs TFET with a heavily doped pocket region in source region near the channel by using three-dimensional atomistic simulations. Compared to conventional P-I-N TFET, the studied TFET has a channel of P-N-I-N. The P-N junction formed in the source region will have a thinner tunneling barrier and introduce an additional tunneling current. Our results demonstrate that high $\texttt{I}_{\texttt{on}}$/$\texttt{I}_{\texttt{off}}$ and much smaller subthreshold slopes can be achieved in the pocket doped TFET. The device performance can be modulated by modifying the length of pocket doped region. By enlarging the pocket length resonant effect and negative trans-conductance will be induced and ambipolar character of TFET is suppressed.

The simulated device structure is shown in Fig. \ref{fig1}(a). The device is a double-gate structure with a gate oxide thickness of  1.0 nm and the dielectric constant of ¦Ê = 4 for  $\texttt{SiO}_2$. GNRs with armchair edge are used as the transport material as shown in Fig. \ref{fig1}(b). The width of ribbon is 1.6nm with the index in the transverse direction N=13. The source region is p-doped with the doping density of  0.01 dopant/atom, and the drain region is n-doped with the same doping density. The channel is an intrinsic armchair GNR(AGNR).  Three regions are of the same length 16nm. In order to improve device performance, a n-doped region with the doping density of  0.01 dopant/atom  is proposed in source near to channel.  $\texttt{L}_\texttt{P}$ is used to denote the length of pocket region.

To study the device characteristics in ballistic limit, open boundary schrodinger equation is solved using the non-equilibrium Green's function formalism self-consistently coupled to a 3D Poisson¡¯s equation\cite{17}. The nearest-neighbor tight binding (TB) approximation with a modified pz orbital model is used to describe the Hamiltonian of the device.  In the model, edge-bond relaxation effect is modeled by using different hopping energy on the edges of the ribbon, t = 1.12 $\texttt{t}_0$. By using the method, the band gap of N=13 AGNR is got of 0.86 eV, which is consistent with the ab initio results\cite{18}.

\begin{figure}
\includegraphics[width=3.2in,height=2.6in]{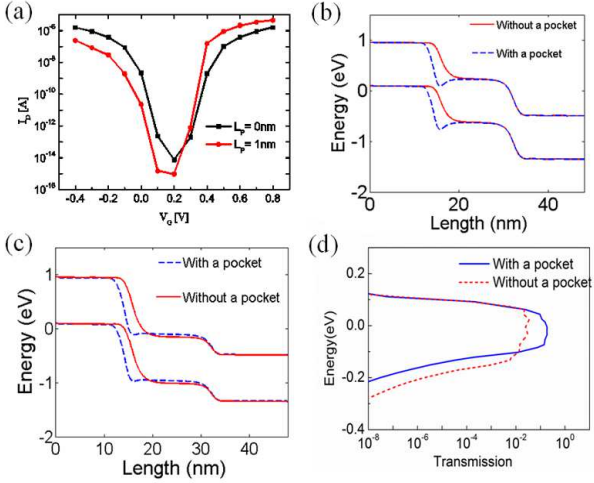}
\caption{(a)Transfer characteristics of  GNRs TFETs with different doping profiles at $\texttt{V}_\texttt{D}$ = 0.4 V. $\texttt{L}_\texttt{P}$ =0 nm and $\texttt{L}_\texttt{P}$  =1nm denote devices without the pocket region and with the pocket region with the length of 1nm, respectively. (b)  and (c) energy band profiles of  the two structure at off state  at $\texttt{V}_\texttt{G}$= 0.2 V and on state at $\texttt{V}_\texttt{G}$=$\texttt{V}_\texttt{D}$, respectively. (d) the compared energy resolved transmission at the on state of  two structure.}
\label{fig2}
\end{figure}

Fig. \ref{fig2}(a) shows transfer characteristics of GNRs TFETs with different doping profiles at a drain voltage $\texttt{V}_\texttt{D}$ = 0.4 V. In conventional P-I-N TFET, the minimum current is achieved at $\texttt{V}_\texttt{G}$ = $\texttt{V}_\texttt{D}$/2, as shown in Fig. 2(a). Fig. \ref{fig2} (a) also shows that the minimum current of P-N-I-N TFET is smaller than that of P-I-N TFET at$\texttt{V}_\texttt{G}$ = $\texttt{V}_\texttt{D}$/2. Furthermore, the on state current of P-N-I-N TFET at $\texttt{V}_\texttt{G}$=$\texttt{V}_\texttt{D}$ is larger than that of P-I-N TFET by one order. As a result, the device performance is improved by using pocket doped source in term of steeper subthreshold sweeps decreasing from  SS = 51.84 mV/dec to SS= 42.80 mV/dec. From the Fig. 2(a), we can also see that the ambipolar character of GNRs TFET is suppressed by using pocket doping.  These results indicate the pocket doping method is an effective way to improve the device performance.

In TFET, there are mainly three aspects which determine the current: tunneling barrier, density of states of injection carrier from source and injection velocity. In the proposed TFETs, we use homo-structure GNRs as channel material and the same doping density in the source region. Thus, the two kinds of TFET have the same density of states and injection velocity from source. The difference of the drain current $\texttt{ I}_{\texttt{D}}$ as a function of gate voltage can be understood from different tunneling barriers. Fig. \ref{fig2} (b) and Fig. \ref{fig2} (c) show energy band diagrams in P-N-I-N TFET and P-I-N TFET in off state and on state, respectively. From Fig. \ref{fig2} (b), we can see that for the structure with a pocket the energy band is pushed to lower energy in the pocket region while the energy  band in the channel keep the same as the structure without the pocket. The phenomenon leads to a larger tunneling barrier in P-N-I-N TFET, which results in smaller current in the off state. Fig. \ref{fig2}(c) shows the energy band profiles of P-N-I-N TFET and P-I-N TFET in the on state at $\texttt{V}_\texttt{G}$= 0.6V. From the figure, we can see that the conduction band in the channel is suppressed below the energy of the valence band in the source region.  Hence there is tunneling current from source valence band to channel conduction band. Due to different doping profiles, the energy bands of the two structures are different.  The energy of channel conduction band of P-N-I-N TFET is higher than that of P-I-N TFET. This means there is narrower tunneling energy windows for P-N-I-N TFET. While, the source pocket doping region introduce a thinner tunneling barrier from source to the channel, and carrier is much easier to tunneling in the channel in P-N-I-N TFET. This is confirmed in the Fig. \ref{fig2} (d), which shows comparison of the energy resolved current spectrum for the two structures at the on state.  It can be seen that P-N-I-N TFET has smaller energy tunneling windows but larger transmission for the reason aforementioned.  As a result, the on state current of P-N-I-N TFET is enhanced by one order compared to P-I-N TFET.

\begin{figure}
\includegraphics[width=3.2in,height=2.6in]{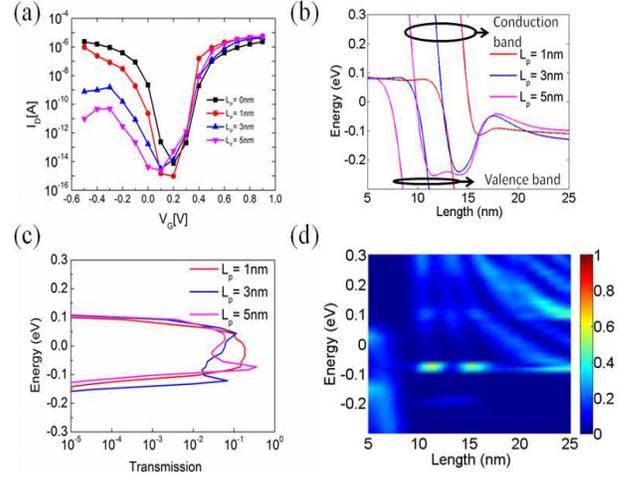}
\caption{ (a) Transfer characteristics of GNRs TFETs with pocket region length of 0nm, 1nm, 3nm and 5 nm. (b) The compared energy band profiles of TFET with pocket region length of 1nm, 3nm and 5 nm at $\texttt{V}_\texttt{G}$=0.4 V. (c) the compared transmission of the three structures. (d)  local density of states of GNRs TFET with pocket region length of 5 nm at $\texttt{V}_\texttt{G}$=0.4 V.}
\label{fig3}
\end{figure}

In order to optimize the device performance of P-N-I-N TFET, we investigate the length effect of pocket doping region as shown in Fig. \ref{fig3}(a). These devices have all the same device parameters except the pocket length.  It can be found that P-N-I-N TFET with 1nm pocket has steepest SS. When pocket length increases from 1nm to 3nm, the on state current at $\texttt{V}_\texttt{G}$ = 0.6 V oscillates.  As aforementioned reason, only tunneling barrier dominates the difference of on sate currents. In order to study the on current oscillation at $\texttt{V}_\texttt{G}$ = 0.6 V  energy potential profiles of devices with different doping pocket length are shown in Fig. \ref{fig3} (b).  It can be found that all these TFETs have a quantum well in the pocket region and 5 nm doping pocket has the deepest well. The well leads to resonant tunneling effect which contributes to the current. In conventional TFET, there is only BTBT current in on state at  $\texttt{V}_\texttt{G}$  = 0.6 V. When there is a pocket doping region, the situation is different. When the length of pocket is small, resonant tunneling isn't formed or make little contribution, such as the case of $\texttt{L}_\texttt{P}$ =1nm.  As long as the length is long enough, resonance appears between pocket region and the channel.  Fig. \ref{fig3} (c) shows transmission as a function of energy. Due to resonance, it can be seen that there are no peak, one peak and two peaks for 1nm, 3nm and 5 nm pocket structures, respectively. From Fig. \ref{fig3} (d), we can see that quantum resonant states exit in the channel of GNRs TFET with pocket region length of 5 nm.  Consequently, resonant effect leads to currents oscillation for different pocket length at  $\texttt{V}_\texttt{G}$ = 0.6 V.

\begin{figure}
\includegraphics[width=3.2in,height=2.6in]{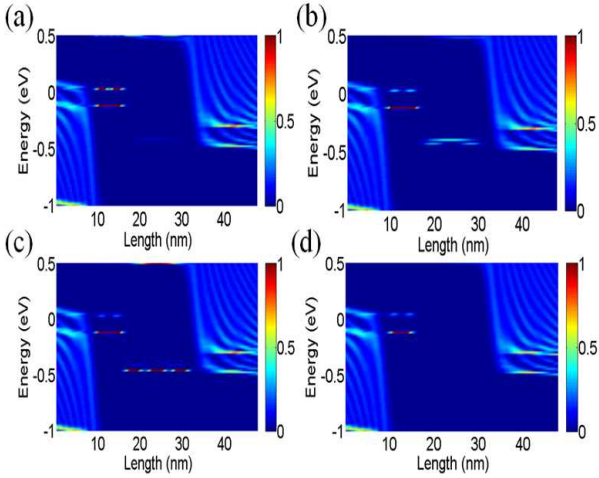}
\caption{ Local density of states of GNRs TFET pocket region length of  5 nm at negative gate voltages of  (a) $\texttt{V}_\texttt{G}$= -0.2 V, (b) $\texttt{V}_\texttt{G}$ = -0.3 V, (c) $\texttt{V}_\texttt{G}$ = -0.4 V and (d) $\texttt{V}_\texttt{G}$ = -0.5 V.}
\label{fig4}
\end{figure}

Even though by modifying the pocket region length, the sub-threshold sweeps can't be improved monotonously. There are two interesting features by increasing the length of pocket region. One is that the ambipolar character of GNRs TFET is dramatically inhabited by increasing the pocket length as shown in Fig. \ref{fig3} (a) . Another interesting feature of the P-N-I-N TFET transfer characteristics is that there is negative trans-conductance in 3nm and 5nm pocket doping P-N-I-N TFET.  This is due to the fact that the states in the pocket region resonant with states in the channel. Fig. \ref{fig4}  shows local density of states of GNRs TFET with 5nm pocket at different negative gate voltages. From Fig. \ref{fig4} (a), it can be easily found that there are resonant states in the pocket region at $\texttt{V}_\texttt{G}$ =-0.2V.   When negative gate voltage is applied to -0.3 V, resonant states appear in the channel as shown in Fig. \ref{fig4} (b) . As gate voltage reaches -0.4 V, resonant effect is more obvious and more resonant states appear in the channel as shown in Fig. \ref{fig4} (c) . However, as the gate voltage keeps deceasing the tunneling barrier becomes thicker and resonant effect in the channel is suppressed. Fig. \ref{fig4} (d) show that resonant states nearly disappear at $\texttt{V}_\texttt{G}$ =-0.5 V. As s result, resonant tunneling current disappears and results in smaller  totoal current, which leads to  negative trans-conductance.

In this paper, simulation study is performed on graphene nanoribbons tunneling field transistor with a pocket doped region in the source. By introducing the n-doped pocket, conventional P-I-N TFET is converted to P-N-I-N structure. By this method£¬ the tunneling barrier from the source to the channel can be reduced, which leads to high on current.  At the same time, due to the pocket TFET has smaller off state current for the thicker direct tunneling barrier from valence band of source to conduction band of drain. As a result, TFET has an improved performance of higher $\texttt{I}_{\texttt{on}}/\texttt{I}_{\texttt{off}}$ and steeper subthreshold slope. Besides, we analyze pocket length on the device performance and find that increasing the length of pocket doping region help to suppress ambipolar character of TFET and leads to resonant effect. The resonance will not only have an impact on on state current but also introduce negative trans-conductance.

This work is supported by NKBRP 2011CBA00604 and 2011ZX02707.

\end{document}